\documentclass[figures,cite,nobm]{epl}
\usepackage{epsf}
\title{Thermoelectric effects in Kondo correlated quantum dots}
\shorttitle{Thermoelectric effects in Kondo dots}
\author{Daniel Boese\inst{1}\thanks{E-mail:
  \email{dboese@tfp.physik.uni-karlsruhe.de}} \and Rosario Fazio\inst{2}} 
\institute{
\inst{1}Institut f\"ur Theoretische Festk\"orperphysik, Universit\"at
  Karlsruhe, D-76128 Karlsruhe, Germany\\
and Forschungszentrum Karlsruhe, Institut f\"ur Nanotechnologie, D-76021
Karlsruhe, Germany\\
\inst{2} Dipartimento Metodologie Fisiche e Chimiche (DMFCI), Universit\`a 
di Catania, Viale A.~Doria~6, I-95125 Catania, Italy\\
and Istituto Nazionale di Fisica della Materia, Unit\`a di
    Catania}
\date{\today}
\begin{document}
\maketitle
\begin{abstract}
In this Letter we study thermoelectric effects in ultra small quantum dots. 
We study the behaviour of the thermopower, Peltier coefficient and 
thermal conductance both in the sequencial tunneling regime and in the 
regime where Kondo correlations develope. Both cases of linear response and 
non-equilibrium induced by strong temperature gradients are considered.
The thermopower is a very sensitive tool to detect Kondo correlations. It changes 
sign both as a function of temperature and temperature gradient. We also 
discuss violations of the Wiedemann-Franz law.
\end{abstract}
{\em Introduction.}
The Kondo effect, studied since several decades in metals~\cite{hewson}, 
has been shown to significantly affect the transport properties through quantum 
dots (QDs). Anticipated theoretically in Refs.~\cite{kondo-theo} it
was  demostrated experimentally in Ref.~\cite{expkondo}. 
Current research in this area aims at exploring features  previously
unaccessible  
in bulk systems. Examples are the study of non-equilibrium 
effects~\cite{meir,koenig96}, 
decoherence~\cite{kaminski}, the phase sensitive  transport~\cite{kondophase}, 
or more exotic models like the singlet-triplet
transition~\cite{pustilnik,eto,giuliano}. 
A yet unexplored problem is the study of thermoelectric effects in
QDs in the  presence of Kondo correlations, which allow to probe for the slope
of the Kondo resonance. 

Transport in the presence of electrical and thermal gradients is a well 
studied phenomenon in bulk systems. More recently it became
possible to explore those effects in nanostructured systems, like 
{\em e.g.}~quantum point contacts~\cite{houten,molenkamp} and 
QDs~\cite{staring}, however, in comparison with electrical transport
thermoelectric effects are still much less studied. Yet they can provide
additional  
information on the kinetics of carriers not contained in the measurement of 
current-voltage characteristics. Moreover they may have interesting technological 
applications using fabricated devices as micro-refrigerators~\cite{nahum}.
In single electron  tunneling devices research  
has focused primarily on the linear thermopower\cite{beenakker,blanter}. More
recently  
the case of small level  spacing  was investigated~\cite{moeller,andreev}. 
Non-equilibrium effects induced by large temperature gradients were not 
studied so far. 

In this Letter we probe the Kondo resonance by studying the thermoelectric power, 
the Peltier effect and the thermal conductance in a QD connected by tunnel 
barriers to two leads. We consider  non-equilibrium situations either due   
to an applied bias voltage or to a temperature gradient.
We make clear predictions which can be verified in experiment. 
The thermopower $S(T)$ changes sign at low temperatures by entering in the Kondo 
regime. If the Kondo resonance is splitted by a magnetic field $S(T)$ shows an 
additional re-entrance. Similar effects are shown when the dot is driven out of 
equilibribium by a strong temperature difference between the electrodes.
We finally discuss the Wiedemann-Franz law and experimental realizations.

{\em Model.}
In ultra-small QDs the level spacing is larger than 
other energy scales like the level broadening or temperature. Moreover
interaction and interference lead to a repulsion of other levels from 
the ``transport-active'' one~\cite{boese}. Transport can be discussed using 
the Hamiltonian for a single spin-degenerate level, tunnel coupled to two leads
$H=H_{\mathrm{res}}+H_{\mathrm{dot}} + H_{\mathrm T}$.
The reservoir part $H_{\mathrm{res}}=\sum_{kr\sigma} \epsilon_{kr\sigma} 
a_{kr\sigma}^\dagger a_{kr\sigma}$, the dot part including the 
interaction $H_{\mathrm{dot}}=\sum_\sigma \epsilon_\sigma c_\sigma^\dagger 
c_\sigma + U n_{\uparrow} n_{\downarrow}$ and the 
tunneling term $H_{\mathrm T}=\sum_{kr\sigma} 
(t^r_{k \sigma} a_{kr\sigma}^\dagger c_{\sigma} + {\mathrm{h.c.}})$. 
In this work we take the charging energy $U \rightarrow \infty$, {\em i.e.}~away
from the symmetric regime.   
The bias voltage shall be applied symmetrically, {\em i.e.}~$\mu_{L/R} 
= \pm q V/2 = \mp eV/2$. The temperature gradient is realized by heating
the left reservoir by $\Delta T$, therefore $T_L=T_R+\Delta T = T + \Delta T$.
We introduce the coupling $\Gamma=\Gamma^L + \Gamma^R$ through
$\Gamma^r (\omega)= 2 \pi \sum_k |t^r_{k \sigma} |^2
\delta(\omega - \epsilon_{k r\sigma})$, which we assume to be independent
of energy and spin. Furthermore we take $\Gamma^r=\Gamma/2$. 
The electrical current and the
heat current at time $\tau$ are  defined respectively
as 
\begin{eqnarray}
        I^r (\tau) &=&i e \sum_{k \sigma} \{ t^r_{k \sigma} \langle (
        a_{kr \sigma}^\dagger c_\sigma ) (\tau) \rangle - \mathrm{h.c.} \}\, ,
\\
        I_Q^r (\tau)&=&i \sum_{k \sigma} \{ (\epsilon_{kr\sigma} 
        - \mu_r) \, t^r_{k \sigma} \langle (
        a_{kr \sigma}^\dagger c_\sigma ) (\tau) \rangle - \mathrm{h.c.} \} \, .
\end{eqnarray}
Note that due to charge conservation the electrical currents in the
left and right reservoirs are the same (up to a sign) while the
heat current due to the factor $\mu_r$ differs by $eV I^r$. This
difference is the rate of entropy production across the QD for an out of 
equilibrium situation. In the following we concentrate on $I_Q^L$.

In the linear response regime the electrical current and the heat 
current   are related to  the bias voltage $V$  and the temperature 
difference $\Delta T$ by the constitutive equations
\begin{equation}
\left(\!\begin{array}{c}I \\ I_Q \end{array}\!\right) =
\left(\begin{array}{cc} G^{(0)} & L^{(0)} 
\\  M^{(0)}  & K^{(0)} \end{array}\right)
\left(\!\!\begin{array}{c} V  \\ \Delta T \end{array}\!\!\right).
\label{eq:thdef1}
\end{equation}
Experimentally however, the current $I$ is  an independent variable
rather than the bias voltage and correspondingly
\begin{equation}
\left(\!\begin{array}{c}V \\ I_Q \end{array}\!\right) =
\left(\begin{array}{cc} R^{(0)} & S^{(0)} 
\\  \Pi^{(0)}  & -\kappa^{(0)} \end{array}\right)
\left(\!\!\begin{array}{c} I  \\ \Delta T \end{array}\!\!\right).
\label{eq:thdef2}
\end{equation}
The Onsager relations tell us that in equilibrium $S^{(0)}=\Pi^{(0)}/T$ and
$L^{(0)}=-M^{(0)}/T$. 

{\em Perturbation theory.}
For temperatures $T>\Gamma$ an expansion in the coupling $\Gamma$ 
is sufficient. We obtain the currents 
for $\Delta \epsilon = 0$ ($\Delta \epsilon $ being the level splitting)
\begin{eqnarray}
I(V,\Delta T) &=& 2 e \Gamma^L \Gamma^R \frac{f_R(\epsilon) - f_L(\epsilon) }{ 
\sum_r \Gamma^r (f_r(\epsilon)+1)} \\
I_Q^r (V, \Delta T) &=& \frac{\epsilon - \mu_r}{-e} I(V, \Delta T) , 
\end{eqnarray}
where $f_r(\epsilon)$ is the Fermi function in reservoir $r$.
Having defined 
$\gamma=2 \Gamma^L \Gamma^R / [ (2 \cosh (\beta \epsilon/2))^2 
\linebreak \sum_r \Gamma^r (f_r(\epsilon)+1)]$ we calculate the thermo-electric
coefficients in linear response to
\begin{equation}
\begin{array}{rclcrclcrcl}
L^{(0)} &=& e \gamma \beta^2 \epsilon & ,& 
M^{(0)} &=& -e \gamma \beta \epsilon  &, & 
K^{(0)} &=& -\gamma \beta^2 \epsilon^2 , \nonumber \\
S^{(0)} &=& -\beta \epsilon / e &, &
\Pi^{(0)} &=& -\epsilon /e &,&
\kappa^{(0)} &=& 0 .
\end{array}
\label{eq:lincoeff}
\end{equation}
The Wiedemann-Franz law is violated due to the strong energy dependence of 
the tunneling rates. 
The thermal conductance $\kappa^{(0)}$ vanishes 
also out of equilibrium (up to exponentially small corrections). The reason is that 
in first order any thermal transport is associated with charge transport.
For a measurement of $\kappa^{(0)}$ however, the electrical current is held 
at zero, and hence $\kappa^{(0)}$ has to vanish. This is not true when higher
orders are included, and therefore $\kappa$ can be used as a primary measurement 
for coherent contributions which manifest themselves in a non-zero value. A similar 
observation holds for 
the Peltier coefficient $\Pi^{(0)}$ which is independent of temperature in 
first order.  Out of equilibrium the expressions become more complicated, 
however, the following analytical relations can be obtained:
\begin{eqnarray}
K(V=0,\Delta T) &=& -\frac{\epsilon}{e} L(V=0,\Delta T) \\
M(V,\Delta T=0) &=& - \left(\frac{\epsilon}{e}+V\right) G(V,\Delta T=0).
\end{eqnarray}
In particular this predicts a sign change of $M(V,\Delta T=0)$ at an 
applied bias voltage of $V=-\epsilon /e$, when the transport switches from
hole to particle like. 

{\em Kondo regime.}
For $T\ll \Gamma$ perturbation theory is no longer sufficient. 
For levels below the Fermi edge spin-fluctuations become 
increasingly important and lead for $T<T_K$ to a new strongly correlated
ground state, in which the dot's spin is completely screened. 
The transition to the Kondo regime is accompanied by the development of
a sharp peak at the Fermi edge in the spectral density. As the 
thermo-electric coefficients are highly sensitive to the structure
(value and slope) of the spectral density near the Fermi edge, they also 
display a strong temperature dependence. 

In order to describe this regime ($T_K<T<\Gamma$) we employ the
non-perturbative resonant tunneling approximation~\cite{koenig96} without
vertex corrections, which is equivalent to the equation of motion
method used in Ref.~\cite{meir}, and gives qualitatively correct 
results for temperatures above the Kondo temperature $T_K$. 
The high-energy cut-off required is chosen to be of Lorentzian-shape 
around $50 \Gamma$ throughout this work. Low $T$ 
calculations for the transport coefficients of the Anderson model 
in equilibrium have been performed using the numerical renormalization
group~\cite{costi}.

\begin{figure} 
  \centerline{\includegraphics[width=8cm]{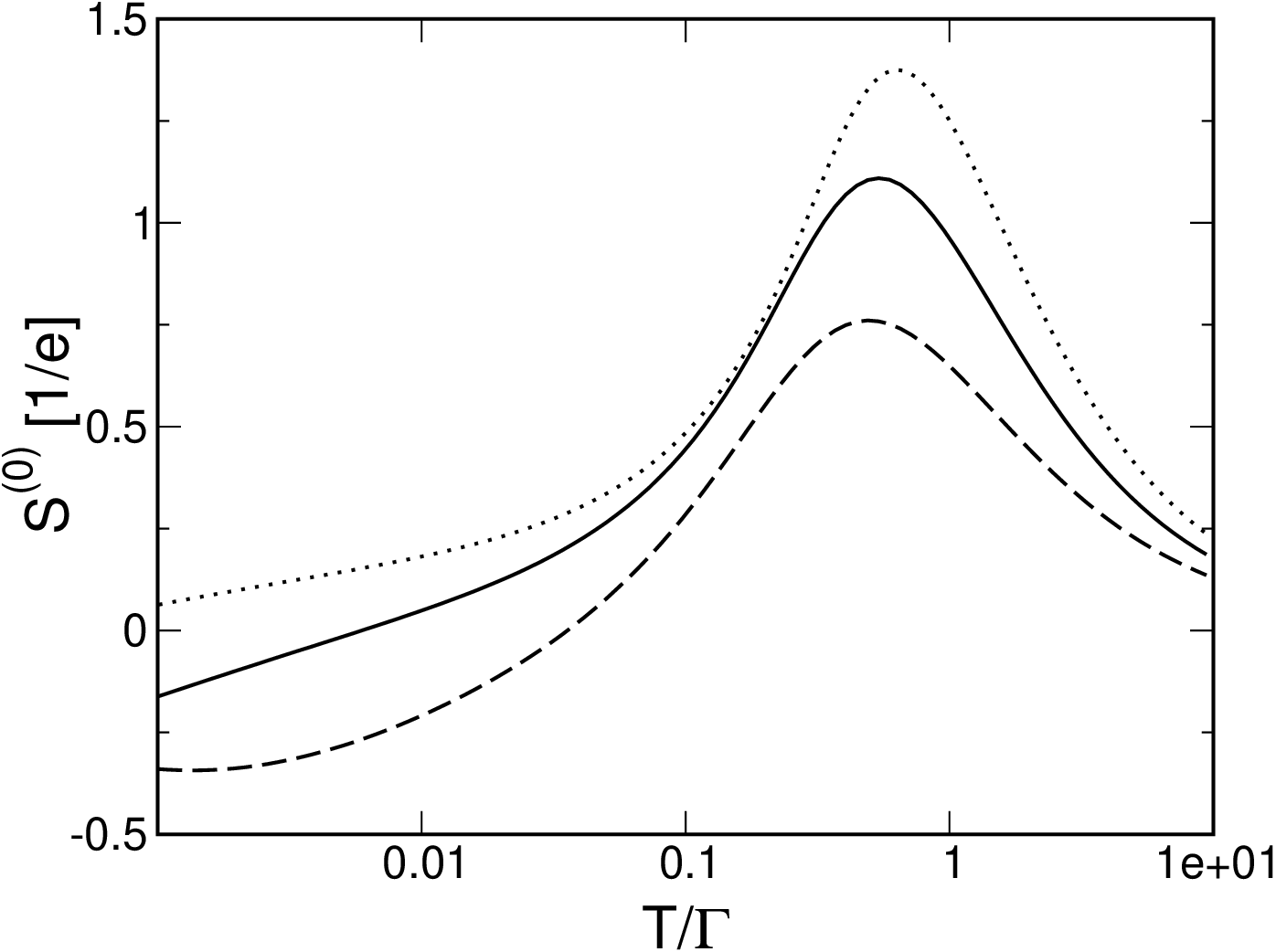}}
        \caption{Linear thermopower $S^{(0)}(T)$ for different level positions
        $\epsilon=-1.5 \Gamma$ (dashed), $-2 \Gamma$ (solid), and 
        $-2.5 \Gamma$ (dotted).}
        \label{fig:linthermo}
\end{figure}

The linear thermopower as a function of temperature is shown in
Fig.~\ref{fig:linthermo} for
three different level positions and hence different Kondo temperatures. 
For $T\cong T_K$ the thermopower approaches a local minimum (not
resolvable in our approximation) from which a logarithmic increase
leads to a broad maximum associated to the single particle excitation
before it reaches the perturbative $1/T$ regime. The important 
point is a change of the sign of $S^{(0)}$ due to the suppression
of the Kondo peak in the spectral density, which should
be easy to measure. 
The change of sign expresses the change from particle like
transport at low temperatures to hole like transport away from the Kondo 
regime. It can also be understood from the fact that the thermopower
is sensitive to the slope of the spectral density at $E_F$.
For a level below (above) $E_F$ the slope is negative (positive). The
Kondo correlations lead to the development of a narrow peak slightly
above $E_F$, which therefore transforms the negative slope into a 
positive one, hence the change of sign. 
Note that the thermopower does not show universal scaling behavior, {\em i.e.} is
not a function of $T/T_K$. In the Kondo regime it is determined by the potential
scattering term, which breaks the particle hole symmetry making $S^{(0)}$ non
zero and which does not scale.  

\begin{figure} 
 \centerline{\includegraphics[width=8cm]{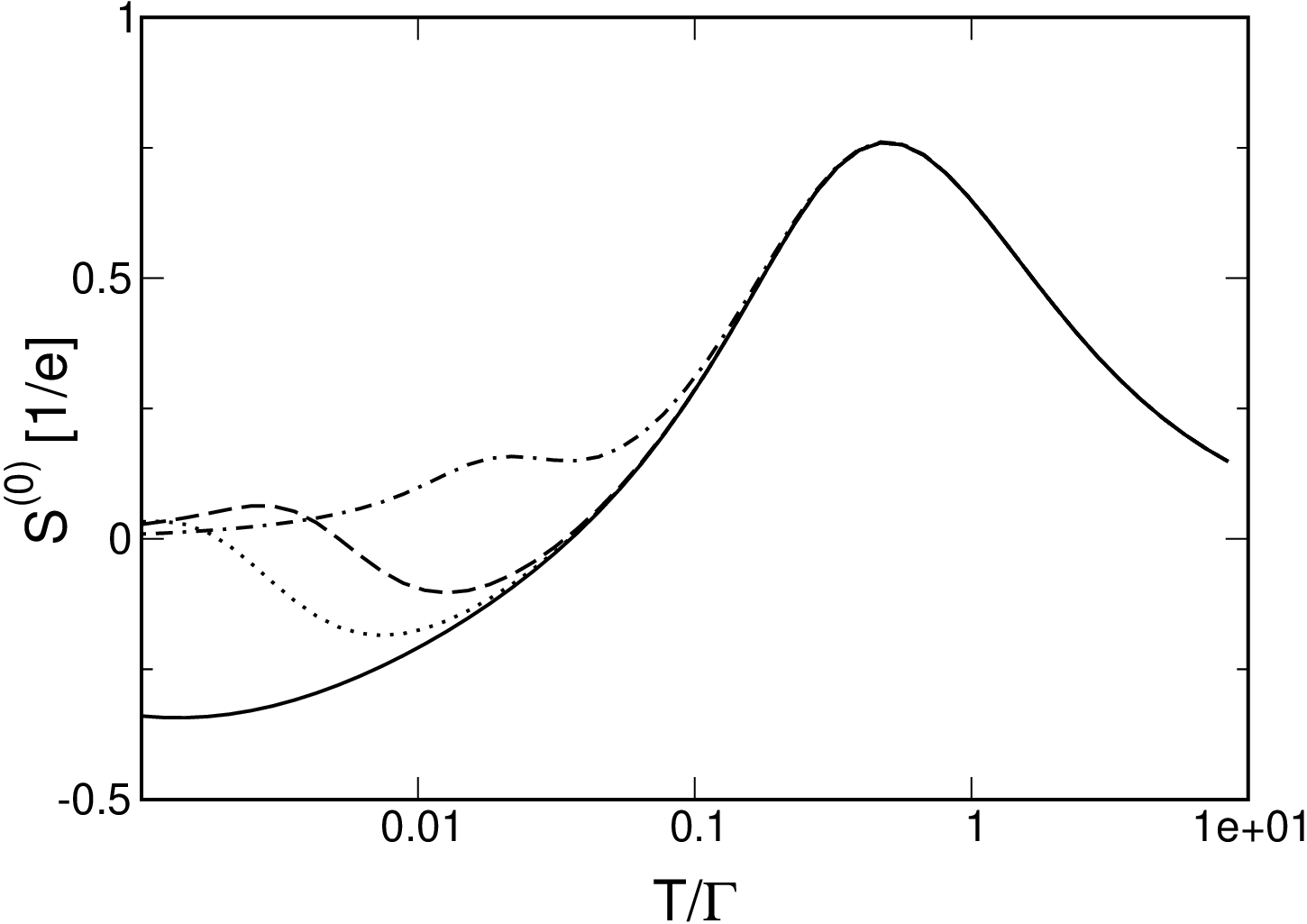}}
        \caption{Linear thermopower $S^{(0)}(T)$ for different level splittings
        $\Delta \epsilon=0$ (solid), $0.01 \Gamma$ (dotted), $0.02 \Gamma$
        (dashed) and $0.1 \Gamma$ (dot-dashed) around $\epsilon=-1.5 \Gamma$.}
        \label{fig:linthermosplit}
\end{figure}

The introduction of a Zeeman splitting can lead to additional structure
as shown in Fig.~\ref{fig:linthermosplit}.  
At temperatures larger than the level splitting $\Delta \epsilon$, $S^{(0)}(T)$  
remains unchanged, around $T\approx \Delta \epsilon$ a local minimum is 
reached resulting in a change of sign at even lower $T$. This can be 
understood from the density of states where for $T>\Delta \epsilon$ the Kondo
peaks starts 
to develop, leading to a negative $S^{(0)}$ at low $T$. Around 
$T\approx \Delta \epsilon$ the peak splits into two, changing the sign
of $S^{(0)}$ and eventually suppressing it when the local minimum
is developed at the Fermi edge.

\begin{figure} 
  \centerline{\includegraphics[width=8cm]{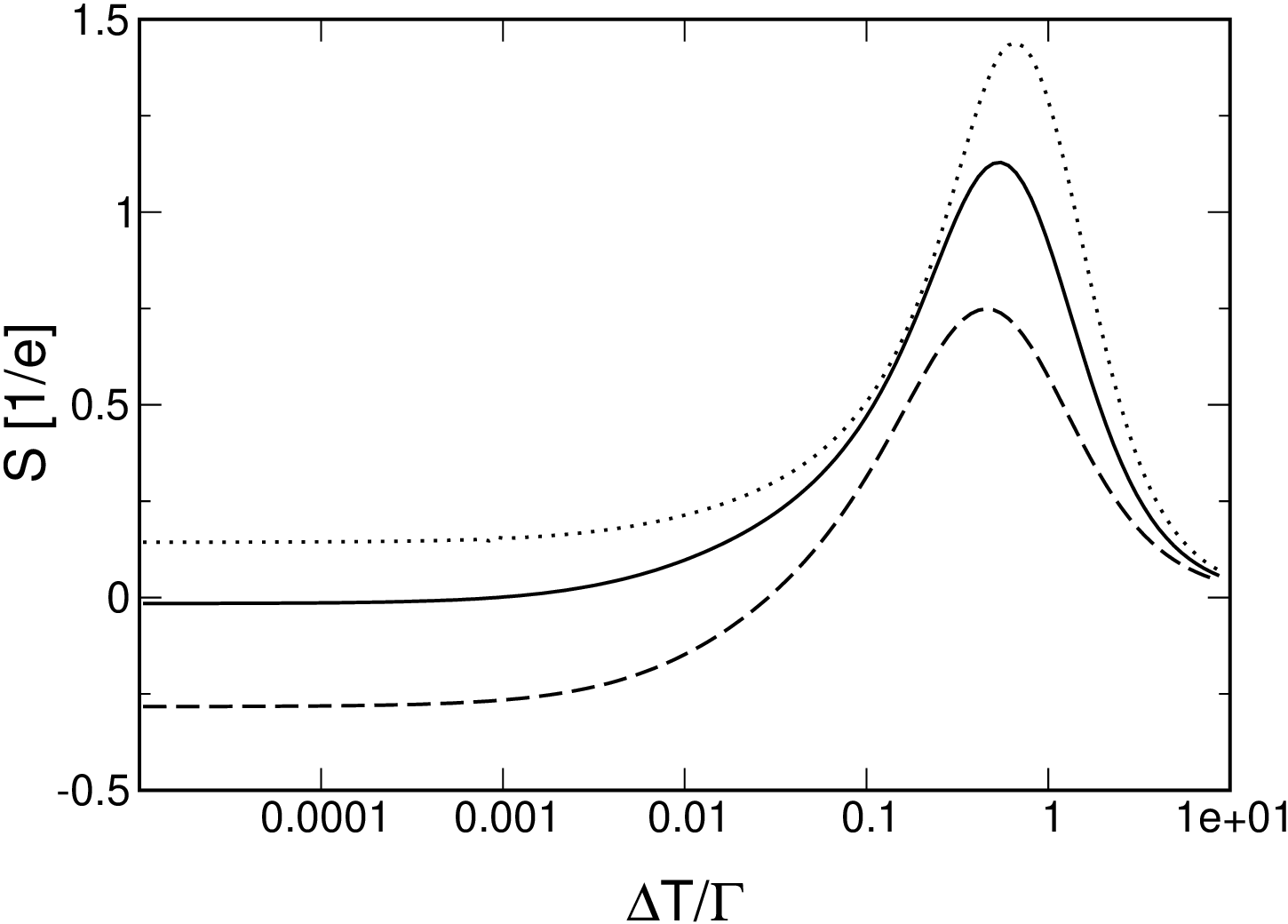}}
        \caption{Differential thermopower $S(T)=\partial V/\partial \Delta T$ 
        for different level positions 
        $\epsilon=-1.5 \Gamma$ (dashed), $-2 \Gamma$ (solid), and 
        $-2.5 \Gamma$ (dotted) at $T=0.005 \Gamma$}
        \label{fig:diffthermo}
\end{figure}

The Seebeck effect is related to the voltage drop which is established through the 
device due to a temperature gradient. The thermopower, given by the 
ratio $V/\Delta T$ is a measure of this effect. Upon increasing $\Delta T$
non-linear effects may appear. We investigate these non-linearities by 
evaluating the voltage $V_t$ at which no current is flowing at a given 
value of  $\Delta T$.
For convenience we consider the differential thermopower 
$S(\Delta T)= \partial V_t / \partial \Delta T |_{I=0}$ and show the results 
in Fig.\ref{fig:diffthermo}. Also $S(\Delta T)$ displays
a change of sign. After departing from the linear regime, characterized by $\Delta T <$
max$\{T_K, T\}$, the previously unaffected Kondo peak is destroyed,
resulting in a logarithmic dependence. Eventually at an energy scale set by
the level position $|\epsilon|$ the single particle peak is reached, where the
physics is dominated by strong charge fluctuations.  

\begin{figure} 
  \centerline{\includegraphics[width=8cm]{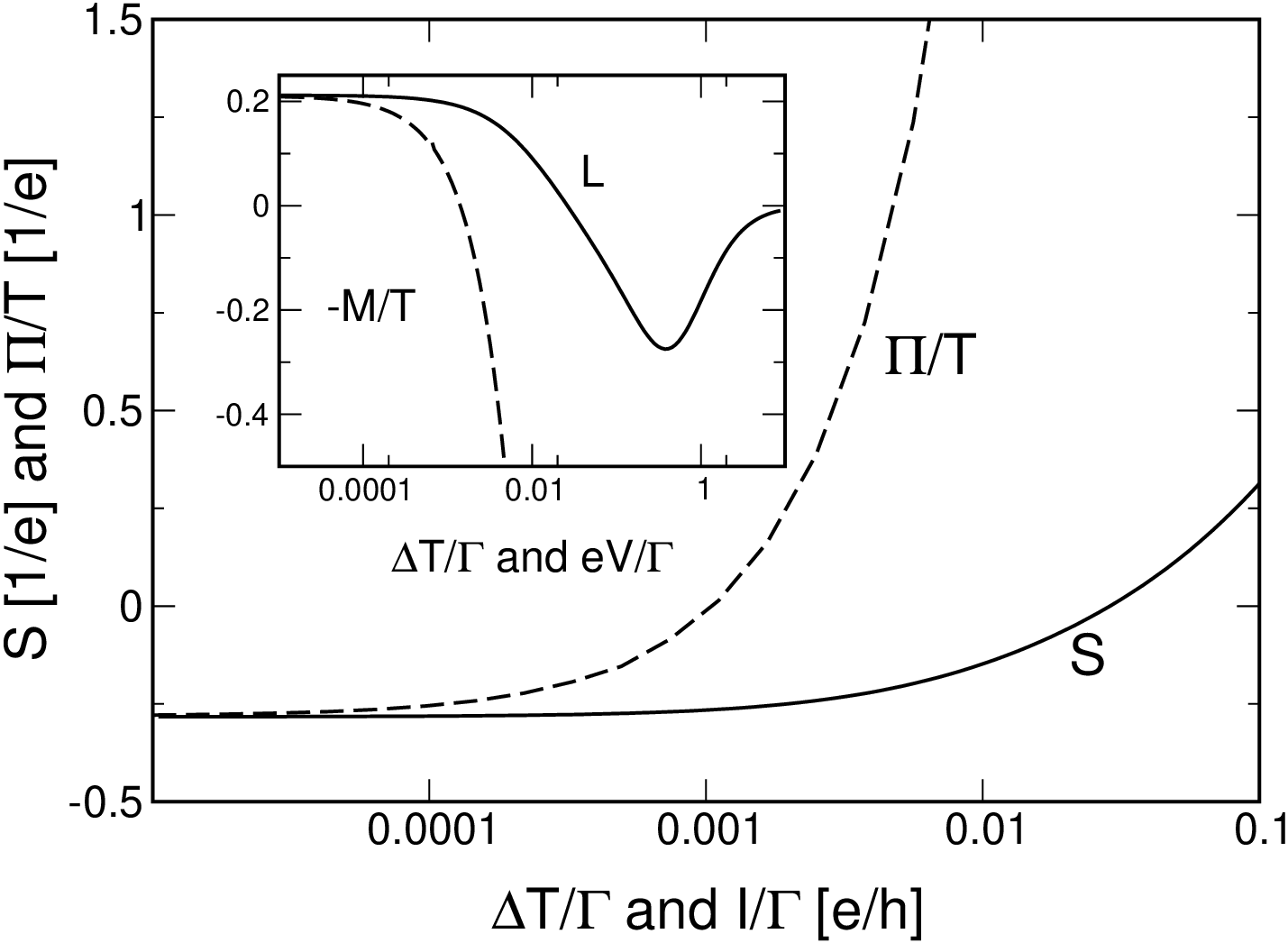}}
        \caption{Differential Seebeck and Peltier coefficient vs.\ their
          driving force. $T=0.005\Gamma$, and $\epsilon=-1.5\Gamma$.
        Inset: Differential $L$ and $-M/T$ in units of $e/h$}
        \label{fig:diffonsagerSpi}
\end{figure}

The Peltier effect is related to the heat current which flows through the device 
in consequence of an electrical current. In linear response the Onsager relations 
relate $\Pi^{(0)}$ to the thermopower. Out of equilibrium however, we study the
differential Peltier coefficient defined as $\Pi(I)= \partial I_Q / \partial I
|_{\Delta  T=0}$. In Fig.~\ref{fig:diffonsagerSpi} we compare 
the pairs of coefficients ($S$, $\Pi/T$) and ($L$, $-M/T$) which 
split upon leaving the linear regime, break the Onsager symmetries, and in
addition can change the sign. 

\begin{figure} 
  \centerline{\includegraphics[width=8cm]{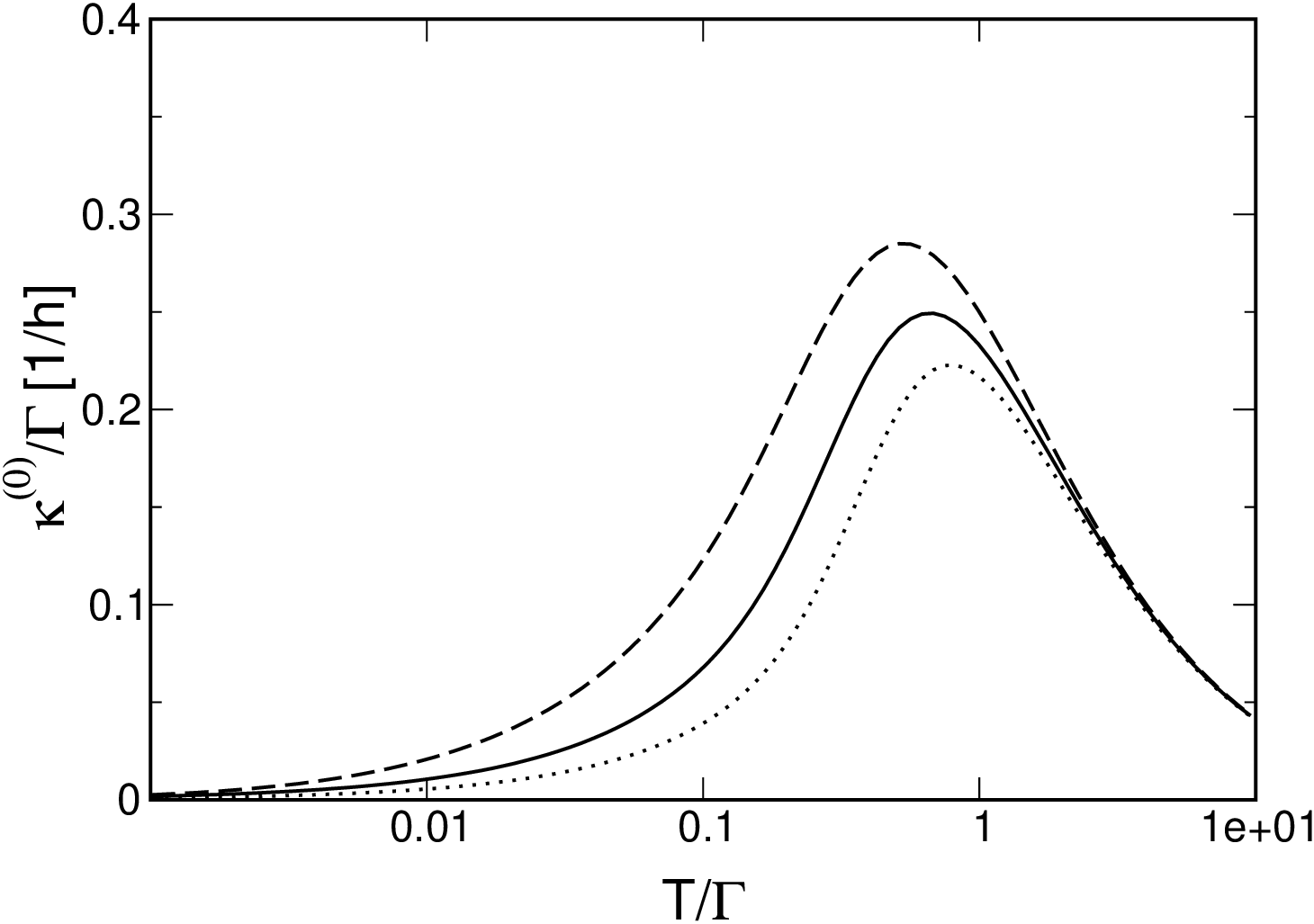}}
        \caption{Linear thermal conductance $\kappa^{(0)}$ for different level 
        positions 
        $\epsilon=-1.5 \Gamma$ (dashed), $-2 \Gamma$ (solid), and 
        $-2.5 \Gamma$ (dotted).}
        \label{fig:linkappa}
\end{figure}

The thermal conductance $\kappa$ is not suitable to measure the
Kondo correlations. As coherent processes measured by $\kappa$ are
quadratically weighted by the energy, contributions close to the 
Fermi edge are negligible. However, we observe a peak related to the
single particle excitation at higher temperatures, as shown in 
Fig.~\ref{fig:linkappa}. The deviations from first order perturbation 
theory, even at higher temperatures, are evident and emphasize the 
importance of higher order processes.

The Wiedemann-Franz law relates thermal and electrical transport 
in metals by the relation $\kappa^{(0)}=L_0 T G^{(0)}$, with the Lorentz number 
$L_0=\pi^2/3 e^2$. In QDs transport occurs differently and
consequently the Wiedemann-Franz law does not hold in general. At 
high temperatures transport is dominated by sequential tunneling events
which leads to a suppression of $\kappa^{(0)}$ relative to $G^{(0)}$. At low 
temperatures however, the Kondo correlations create an effective Fermi liquid 
state for which the Wiedemann-Franz law holds again. The crossover shows the
logarithmic temperature dependence characteristic for the onset of Kondo
correlations modified by a the peak seen in Fig.~\ref{fig:linkappa}.

Finally we remark on the experimental check of our results. 
The definition of the Seebeck and Peltier coefficients imply current
control, which due to non-perfect measurement devices and finite impedances of
the external circuit might not be perfectly achieved\cite{amman}. Although
this could render the observability of vanishing $\kappa$, the conclusions
regarding the Kondo correlations remain valid. However, using the Onsager
relations as probe for linear response will no longer be well defined.

{\em Summary.} We have studied electrical and heat transport through
voltage and temperature biased QDs. Using
perturbation theory we found analytic expressions for all coefficients
in first order. We found that the measurement of the thermal 
conductance $\kappa$ allows to directly probe higher-order coherent contributions,
and that the Wiedemann-Franz law is strongly violated at higher 
temperatures. We have shown that the development
of Kondo correlations leads to logarithmic dependencies in the
thermoelectric coefficients, temperature dependent changes of sign, 
and made clear predictions for experimental observation. 
\acknowledgments
We would like to thank Ya.~M.~Blanter, A.~Rosch, H.~Schoeller and J.~Siewert
for valuable 
discussion. This work is supported by the  
DFG as part of the Graduiertenkolleg ''Kollektive Ph\"anomene im
Festk\"orper'' and ``SFB 195'' (D.B.), and by the EU TMR network
``Dynamics of Nanostructures''. D.B. wishes to thank the group in Catania for
their hospitality during his stay in Catania.

\end{document}